\documentclass[aps,prb,10pt,twocolumn,showpacs,preprintnumbers,amsmath,amssymb]{revtex4-1}

\usepackage{float}
\usepackage{graphicx}
\usepackage{tikz}
\usepackage{dcolumn}
\usepackage{color}
\usepackage{amsmath}

\begin{document}
\title{Nonordinary  criticality at the edges of  planar spin-1 Heisenberg antiferromagnets }

\author{Lukas Weber}
\affiliation{Institut f\"ur Theoretische Festk\"orperphysik, JARA-FIT and JARA-HPC, RWTH Aachen University, 52056 Aachen, Germany}

\author{Stefan Wessel}
\affiliation{Institut f\"ur Theoretische Festk\"orperphysik, JARA-FIT and JARA-HPC, RWTH Aachen University, 52056 Aachen, Germany}

\date{\today}

\begin{abstract}
Dangling edge spins of dimerized two-dimensional spin-1  Heisenberg antiferromagnets are shown to exhibit  nonordinary quantum critical correlations, 
akin to the  scaling behavior observed in  recently explored  spin-1/2 systems. Based on large-scale quantum Monte Carlo simulations, we observe  remarkable similarities  between these two cases and examine the crossover to the fundamentally distinct 
behavior in the one-dimensional limit of strongly coupled edge spins. 
We  complement our  numerical analysis by a cluster mean-field theory that encompasses the qualitatively similar behavior for the spin-1 and the spin-1/2 cases and its dependence on the spatial edge-spin configuration in a generic way.
\end{abstract}

\maketitle

\section{Introduction}
\label{sec_intro}
Many aspects of quantum critical magnets  can be described in terms of  an effective classical field theory. This applies, in particular, to quantum critical points of unfrustrated quantum antiferromagnets, for which the quantum-to-classical mapping provides a description of  the quantum critical properties of a $d$-dimensional quantum system  in terms of a ($d+1$)-dimensional  classical $\phi^4$ field theory~\cite{Sachdev11}. For a SU(2)-symmetric system, the effective field theory 
contains a three-component $\phi$ field with an O(3)-symmetric action, which  also describes, e.g.,  the thermal criticality of classical  Heisenberg ferromagnets. 

An interesting twist to this relationship is provided by considering surface critical phenomena in quantum magnets. Whereas the field of classical surface criticality is  rather mature, and a systematic  theory based on the renormalization group  has been developed early on (see, e.g., Ref.~\onlinecite{Diehl86} for an extended review), recent work~\cite{Zhang17,Ding18,Weber18} uncovered surprises when it comes to applying these results to a corresponding low-dimensional quantum magnetic system: Most striking  in this respect is the observation that several two-dimensional unfrustrated quantum critical magnets may exhibit values of the algebraic scaling exponents at appropriately prepared edges that are not observed at surfaces of  the corresponding
three-dimensional classical Heisenberg model. In particular, for the O(3)-symmetric  case, the Mermin-Wagner theorem forbids the presence of a finite-temperature surface transition above the bulk critical temperature~\cite{Mermin66}. In effect, the classical surface exhibits algebraic correlations only at the bulk's critical temperature, defining the  bulk-induced,  ordinary surface universality class. 

It was, indeed, observed recently  in various unbiased numerical studies that two-dimensional SU(2)-invariant Heisenberg antiferromagnets exhibit algebraic correlations at the edges of a quantum critical bulk that are   
 in accord with the scaling exponents of the ordinary surface universality class~\cite{Zhang17,Ding18,Weber18}.
However, this is not the only possibility: In fact, it was  found that
  such systems  exhibit a remarkably distinct nonordinary  power-law scaling behavior for appropriately constructed edge-spin configurations, characterized by so-called dangling edge spins~\cite{Zhang17,Ding18,Weber18}. 

\begin{figure}[t]
\includegraphics[width=0.5\columnwidth]{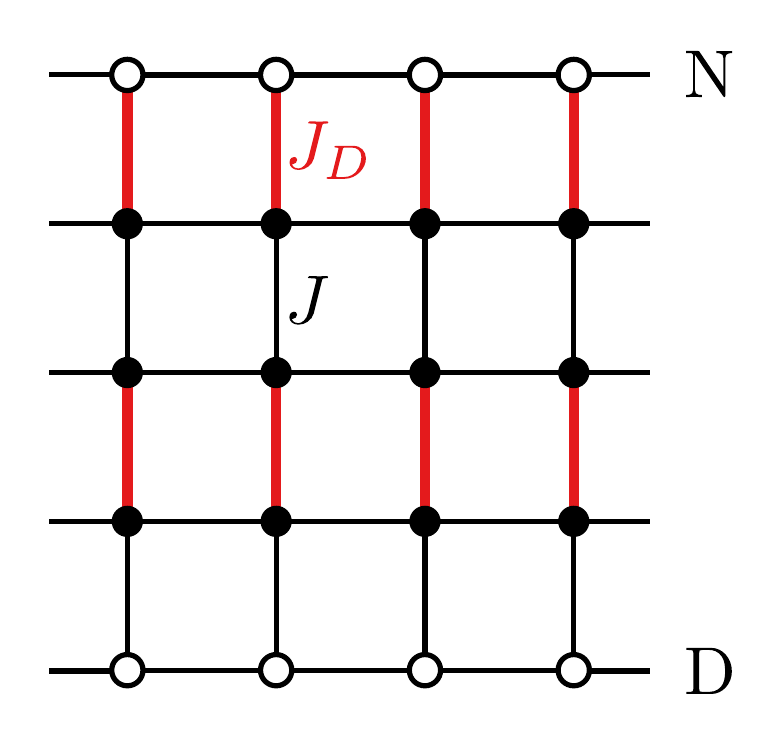}
\caption{  Columnar-dimer lattice with nondangling (N) edge spins (top edge) and dangling (D) edge spins (bottom edge). Solid (open) circles show bulk (edge) spins, and  thick red (thin black) lines denote intra- (inter) dimer couplings $J_D$ ($J$). }
\label{fig_lattice}
\end{figure}

A simple model that allows us to illustrate  this scenario is shown in Fig.~\ref{fig_lattice}: Here, we consider spin-$S$ degrees of freedom located on the sites of a square lattice with SU(2)-invariant Heisenberg exchange interactions along the nearest-neighbor bonds.
The exchange constants are arranged, such as to  form a columnar system  of coupled spin dimers. Denoting the (stronger) intradimer coupling as $J_D$ and the interdimer coupling $J$, this system for $S=1/2$ is well known to exhibit a quantum critical point at a value of $J/J_D= 0.523\,37(3)$~\cite{Matsumoto01,Wenzel08}, which separates a phase with  antiferromagnetic order  from the quantum disordered regime of strong dimer coupling $J_D$.
In addition,  Fig.~\ref{fig_lattice} illustrates two different kinds of edges: The edge spins at the top edge are each connected to another spin by a strong dimer coupling $J_D$, whereas for the configuration shown at the bottom, the edge spins are in that respect missing their strong-coupling partner. We denote these two possibilities as N and D edge spins, respectively. 

As detailed in Refs.~\onlinecite{Zhang17,Ding18,Weber18} for the spin-1/2 case, the edge spins  exhibit  algebraic power-law correlations for both kinds of edges if the ratio $J/J_D$ is tuned to the bulk critical value. 
However, the dangling edge-spin configuration  exhibits nonordinary values of the corresponding critical exponents, in contrast to the nondangling case for which the power-law exponents are in good accord with  established values for the ordinary surface transition of the three-dimensional O(3) theory. Moreover, the differences between the dangling and the nondangling exponents  are  rather substantial, and in some cases  involve differences even in the overall sign. 
Such a distinction was also observed  in other spin-1/2 antiferromagnets for which a corresponding formation of edges with dangling or nondangling edge spins can be realized. Moreover, the nonordinary scaling exponents for the dangling cases take on values that  compare well among the various considered models, even though weak variations in the reported numerical values were  observed. This may be taken as an indication for a distinct universality class underlying this nonordinary surface criticality. 

It was, furthermore, noted in Ref.~\onlinecite{Ding18} that the numerical values of the nonordinary exponents are  similar to those obtained by a renormalization group calculation in second-order $\epsilon$ expansion (around four dimensions)  for the special surface transition  of the classical O($N$) model---after a naive extrapolation to $\epsilon=1$ and setting $N=3$. 
The fact that the classical three-dimensional Heisenberg ferromagnet, corresponding to $N=3$, does not feature this special transition,  was  argued to be due to the proliferation of topological defects in the effective $\phi$ field configurations, which are not accounted for by the renormalization-group approach~\cite{Ding18}. It was, thus, suggested in Ref.~\onlinecite{Ding18}, that  
the topological $\theta$ term in the effective field theory of the spin-1/2 antiferromagnetic Heisenberg chain leads to the suppression of these topological defects and thereby stabilizes the special surface transition for the case of the  spin-1/2 quantum model. 
As is well known, this topological $\theta$ term arises from the spin Berry phase and is associated with the
gapless, quantum critical ground state of the spin-1/2 Heisenberg chain. 
This is in marked contrast to the case of the spin-1 (or any other integer-$S$) Heisenberg chain, which is described by the standard nonlinear $\sigma$ model without a 
topological $\theta$ term and which is, instead, characterized  by an exponential decay of the spin-correlation function and a finite magnetic excitation gap~\cite{Haldane81,Haldane83a,Haldane83b,Affleck85a,Affleck85b,Haldane85}. 

In view of these arguments, it is  not clear, whether  the nonordinary edge criticality  observed for  dangling edge spins of quantum critical spin-1/2  magnets may in fact also appear in the spin-1 case or whether, in this case, instead, both nondangling as well as dangling edge spins  exhibit ordinary  critical exponents. In this paper, we address this question by means of  unbiased quantum Monte Carlo (QMC) simulations for the specific case of the columnar-dimer lattice in Fig.~\ref{fig_lattice}. We provide clear evidence for the emergence of nonordinary exponents also in the case of dangling spin-1 edges, whereas for the nondangling case, we recover the ordinary scaling exponents from the classical theory. Furthermore, we examine the crossover from the  edge-spin system to the strongly coupled edge-spin chain limit for which the distinction between the half-integer vs integer spin-$S$ case is eventually recovered. Finally, we provide a  MF theory in terms of the dimer units of the columnar-dimer lattice as a simple  approximative analytical treatment of these quantum critical systems. 
The further layout of this paper is as follows: In Sec.~\ref{sec_model}, we introduce the model system in more detail as well as the QMC approach that we used for our numerical analysis. The results of our numerical studies are presented Sec.~\ref{sec_res}, and the cluster mean-field (MF) theory is introduced in Sec.~\ref{sec_mf}. Final conclusions are drawn in Sec.~\ref{sec_conclusions}. 

\section{Model and QMC Method}
\label{sec_model}
In the following, 
we examine the $S=1$ columnar-dimer antiferromagnet, described by the Hamiltonian
\begin{equation}
H=J \sum_{\langle i,j \rangle} \mathbf{S}_i \cdot  \mathbf{S}_j +  J_D \sum_{\langle i,j \rangle_D} \mathbf{S}_i \cdot  \mathbf{S}_j ,
\label{eq_hamiltonian}
\end{equation}
where the first summation extends over the bonds of strength $J$, coupling neighboring spin from different dimers, and the second 
summation extends over the dimer bonds of strength $J_D$, cf. Fig.~\ref{fig_lattice}. 
To examine this model by an unbiased numerical means, we used the stochastic series expansion QMC approach with directed loop updates~\cite{Sandvik91,Sandvik99,Henelius00} for all the numerical simulations reported in this paper. 
Similar to the spin-1/2 case, 
at zero temperature $T=0$, this system undergoes a continuous quantum phase transition
separating  antiferromagnetic order from the  quantum disordered phase of strong dimer coupling at a  critical coupling ratio,
which can be extracted as approximately $J/J_D\approx 0.19$ from Fig.~11 of Ref.~\onlinecite{Matsumoto01}.
We are not aware of a  precision estimate of the location of the critical point, which is required for our purpose and, thus, performed a standard finite-size scaling analysis to determine this bulk quantity, based on QMC simulations of finite systems with periodic boundary conditions, cf. Appendix~\ref{appa}.  
From this analysis, we obtained a value of the critical interdimer coupling of $J/J_D = 0.189\,20(2)$ for the $S=1$ case.  

When open boundary conditions are applied along one of the lattice  directions, as shown in Fig.~\ref{fig_lattice}, there may be either interdimer $J$  bonds or  $J_D$ dimer bonds terminating on the surface. This gives rise to either dangling  or nondangling edge spins, respectively, cf.~ Fig.~\ref{fig_lattice}. In order to examine the properties of these edge-spin subsystems at the bulk quantum critical point, we simulated systems of $L\times L$ two-spin dimer unit cells (with a total of $2 L^2$ spins), scaling the temperature as $T=J_D/(2L)$ since the dynamical critical exponent for the bulk transition is $z=1$ in order to probe the ground-state properties. In the following section, we present the numerical results from these QMC simulations.

\section{QMC Results}
\label{sec_res}
Here, we are interested in the behavior of the edge-spin system when the bulk is tuned to the quantum critical point, 
$J/J_D = 0.18920(2)$. In this case, in addition to the bulk, the edge spins also exhibit 
power-law behavior, featuring distinct critical exponents. 
To extract these, we measured the spin-spin 
correlations  $\langle S^z_i S^z_j \rangle$ among  two  edge spins $i,j$ at a distance $r$ parallel to  the edge, denoted $C_\parallel(r)$, as well as the correlations  $C_\perp(r)$
between an edge spin $i$  and an equivalent bulk spin $j$ (with respect to the unit cell) at a distance $r$ perpendicular to the edge.
We consider, in particular, the value of these correlations at the maximum distance $L/2$ on the finite clusters,
 $C_\parallel(L/2)$ and $C_\perp(L/2)$. Monitoring the $L$-dependence of these quantities allows us to monitor the corresponding correlations as a function of distance, based on QMC data from the longest accessible distance on each finite system. 
In addition, we measured
the staggered susceptibility $\chi_s$ of the edge-spin subsystem using the Kubo integral~\cite{Sandvik91}, $\chi_s=\frac{1}{L}\int_0^\beta d\tau \: \langle M_s(\tau) M_s(0)\rangle$ in terms of the staggered edge moment $M_s=\sum'_i \epsilon_i S^z_i$. Here, the summation is restricted to the edge spins (with $\epsilon_i=\pm 1$, depending on the sublattice to which site $i$ belongs). 

\subsection{Scaling behavior}
The observables that we  introduced above are related to the surface critical exponents  $\eta_\parallel$, $\eta_\perp$,  and $y_{h_1}$ via the scaling laws~\cite{Zhang17,Ding18,Weber18}
(for the spatial dimension  $d=2$ and the dynamical critical exponent $z=1$), 
\begin{align}
\label{eq_scaling}
 C_\parallel(L/2) &\sim L^{-(d+z-2+\eta_\parallel)},\\
 C_\perp(L/2) &\sim L^{-(d+z-2+\eta_\perp)},\\
 \chi_s &\sim L^{-(d+z-1-2y_{h_1})}.
\end{align}
In correspondence to classical surface criticality, one, furthermore, 
expects the following scaling relations~\cite{Diehl86} to hold among these exponents~\cite{Zhang17,Ding18,Weber18}: 
\begin{equation}\label{eq_scalingrelations}
\eta_\parallel=3-2y_{h_1}, \quad 2\eta_\perp=\eta_\parallel+\eta.
\end{equation}
Here,  $\eta$ is the anomalous dimension at the bulk transition with
 $\eta=0.0357(13)$ for the three-dimensional O(3) universality class~\cite{Campostrini02}. 

\begin{figure}[t]
\includegraphics{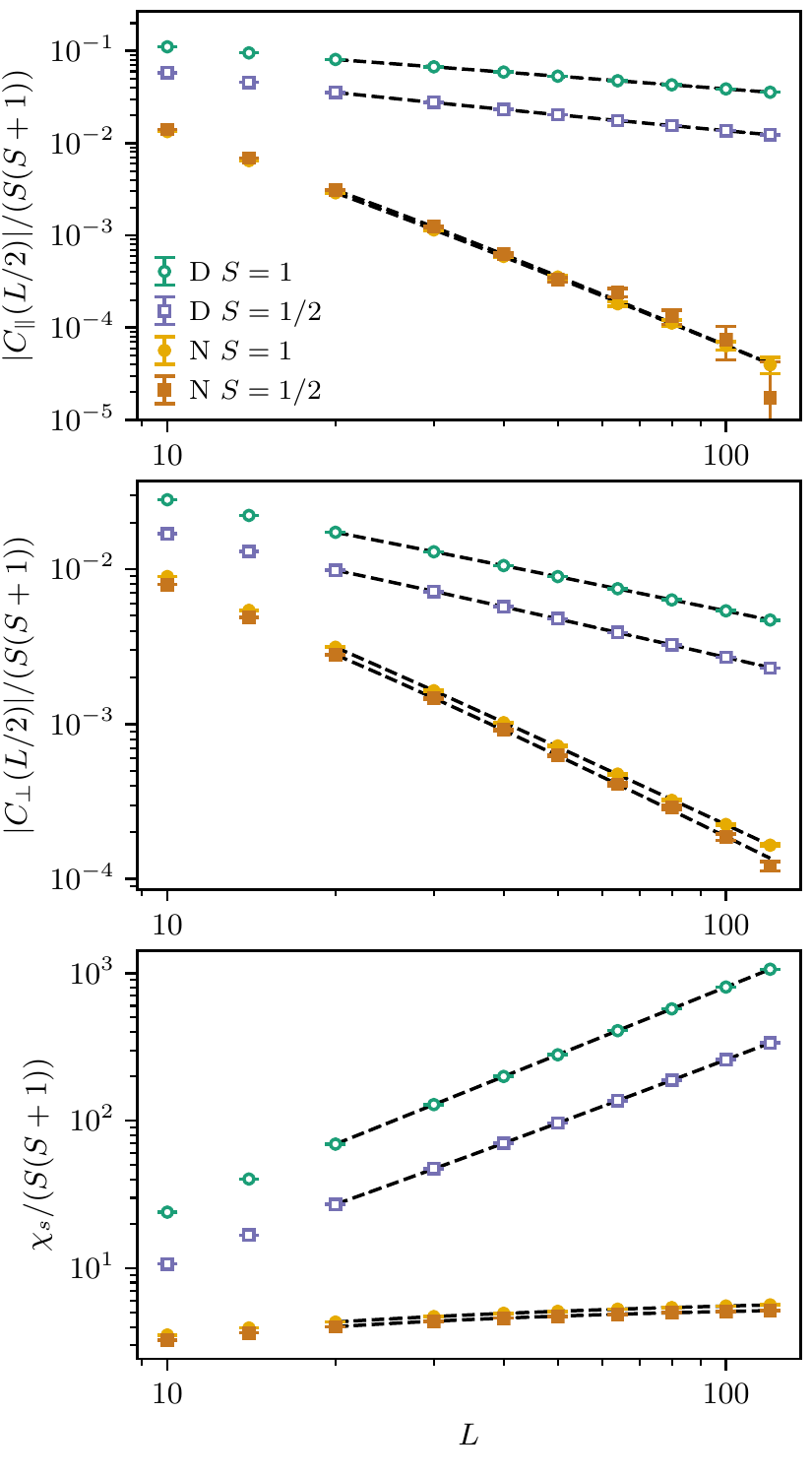}
\caption{  Correlations parallel, $C_\parallel(L/2)$ (top panel), perpendicular,  $C_\perp(L/2)$ (middle panel), and the staggered susceptibility $\chi_s$ (bottom panel), as  functions of $L$ for $S=1$, and $S=1/2$ for the two different edge configurations. The dashed lines indicate the fits as described in the  text.}
\label{fig_critexp}
\end{figure}

Results for the linear system-size dependence of the three quantities are shown for both the D and N cases  in Fig.~\ref{fig_critexp} where, for comparison, we also show data for the $S=1/2$ case on the same finite lattices at the corresponding quantum critical coupling strength of $J/J_D = 0.52337(3)$~\cite{Matsumoto01,Wenzel08}. We observe a rather similar distinction between the dangling and the nondangling cases for both values of the quantum spin $S$ in terms of the $L$ dependence of the correlation functions. The  difference between the spin-1/2 and the spin-1 cases mainly concerns the overall magnitude of the correlations, which are enhanced for the spin-1 case, reflecting  its reduced quantum character. 
A corresponding factor of  $S(S+1)$ has been accounted for in the normalization of Fig.~\ref{fig_critexp}.
We observe that in the  nondangling case, these normalized correlations for $S=1/2$ and $S=1$ are particularly  close.
 
Concerning the  scaling of these quantities with the linear system size $L$, both spin values exhibit very similar slopes in Fig.~\ref{fig_critexp} for both the dangling and the nondangling cases, which quantify the scaling exponents according to Eq.~(\ref{eq_scaling}).
For the estimation of these exponents from the finite-size data, residual finite-size corrections to scaling have to be taken into account. 
Following the  finite-size analysis for the spin-1/2 case~\cite{Zhang17,Ding18,Weber18},  we used the following general scaling ansatz,
\begin{equation}
f(L) = a L^{b} (1 + c_1 L^{-\omega}) + c_{ns}
\end{equation}
for each of the three considered quantities  $C_\parallel(L/2)$, $C_\perp(L/2)$, and   $\chi_s$ individually, 
where $a$, $b$, $c_1$, and $c_{ns}$ are fitting parameters. In particular, the exponent $b$  directly relates to the 
surface critical exponents, according to Eq.~(\ref{eq_scaling}).
The $L^{-\omega}$ term takes into account the leading correction to scaling (in practice, we fix $\omega=1$ as in previous work~\cite{Zhang17,Ding18,Weber18}).
We included this correction term whenever its exclusion did not provide an acceptable fit of the finite-size data.   
Finally, the constant term $c_{ns}$ is included in the fitting ansatz for $\chi_s$ for the nondangling configuration since in that case the nonsingular background  provides the leading contribution to the susceptibility~\cite{Zhang17,Ding18,Weber18}. 
 Whenever $c_1$ and $c_{ns}$ were not included as fit parameters, they were fixed to 0,   see also Table~\ref{table_fitparams} for 
 a summary of this fitting procedure. Finally, we note that all fits were restricted to system sizes $L\geq 20$.
\begingroup
\squeezetable
\begin{table}
\begin{ruledtabular}
\begin{tabular}{clcc}
Exponent & Configuration & $c_1 L^{-1}$ included & $c_{ns}$ included \\
\hline\hline
$\eta_\parallel$ & Nondangling & No & No \\
 & Dangling & Yes & No \\
\hline
$\eta_\perp$ & Nondangling & Yes & No \\
 & Dangling & Yes & No \\
\hline
$y_{h_1}$ & Nondangling & No & Yes \\
 & Dangling & Yes & No \\
\end{tabular}
\end{ruledtabular}
\caption{Correction terms included in the estimation of the various surface critical exponents.}
\label{table_fitparams}
\end{table}
\endgroup

\begingroup
\squeezetable
\begin{table}
\begin{ruledtabular}
\begin{tabular}{lcccc}
 Configuration & Spin $S$& $\eta_\parallel$ & $\eta_\perp$ & $y_{h_1 }$ \\
\hline
\noalign{\vskip 1mm}
Nondangling   & $1$  &  $1.32(2)$ & $0.70(2)$ & $0.80(1)$\\
&$1/2$ & $1.30(2)$ & $0.69(4)$ & $0.84(1)$\\
                       \hline
Dangling  & $1$    &  $-0.539(6)$ & $-0.25(1)$ & $1.762(3)$\\
& $1/2$ & $-0.50(1)$ & $-0.27(1)$ & $1.740(4)$\\
\end{tabular}
\end{ruledtabular}
\caption{Surface critical exponents $\eta_\parallel$,  $\eta_\perp$, and $y_{h_1 }$ for the dangling and nondangling edges of the columnar-dimer lattice  shown in Fig.~\ref{fig_lattice}. The values for $S=1/2$ are quoted from Ref.~\onlinecite{Weber18}.}
\label{table1}
\end{table}
\endgroup

From this finite-size analysis, we  obtained the 
estimates for the $S=1$ surface critical exponents shown in Table~\ref{table1}, 
which also contains the corresponding values for the $S=1/2$ case, taken from Ref.~\onlinecite{Weber18}.
Based on this quantitative comparison, we  confirm the observation from Fig.~\ref{fig_critexp}, that, for both $S=1/2$ and $S=1$, the dangling edge spins exhibit rather distinct scaling properties as compared to the case of nondangling edge spins.
For the later case, the critical exponents that we obtained for the $S=1$ case are in good  agreement with the standard estimates for the ordinary surface transition of the three-dimensional O(3) model, similar to the previously examined $S=1/2$ case~\cite{Zhang17,Ding18,Weber18}.
Moreover, also in the case of dangling edge spins, the 
estimates of the nonordinary critical exponents are similar for the two considered values of $S$. 
Within the estimated uncertainty, these values are also  compatible with the scaling 
relations in Eq.~(\ref{eq_scalingrelations}).
This shows that also dangling spin-1 edge spins exhibit nonordinary surface criticality of a form that compares well to the $S=1/2$ case in terms of the scaling properties.

\begin{figure}[t]
\includegraphics[width=0.5\columnwidth]{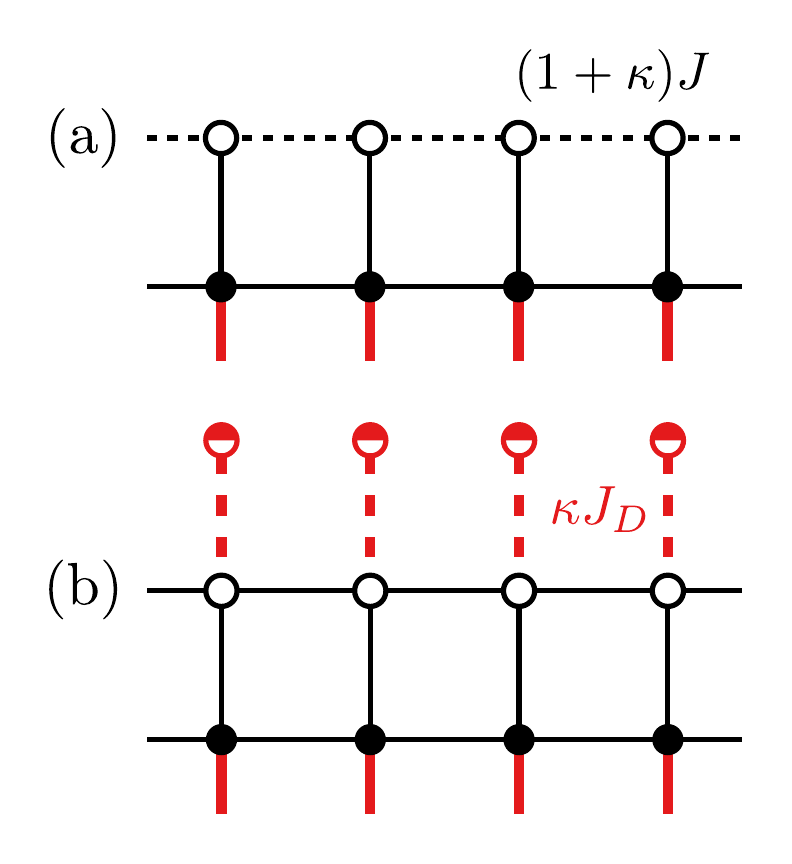}
\caption{  Two different perturbations that we applied to the dangling edge  spins. 
In panel (a), the intra-edge couplings get rescaled by a factor of $(1+\kappa)$, whereas in panel (b), additional spins interact  with the edge spins via a Heisenberg coupling of strength $\kappa J_D$. In both cases, $\kappa=0$ corresponds to the original model. In both panels, the  varied couplings are shown by dashed lines. In panel (b), the half-open circles denote the additional  spins that couple each to an original dangling edge spin. }
\label{fig_perturbations}
\end{figure}

\subsection{Edge perturbations}

\begin{figure}[t]
\includegraphics{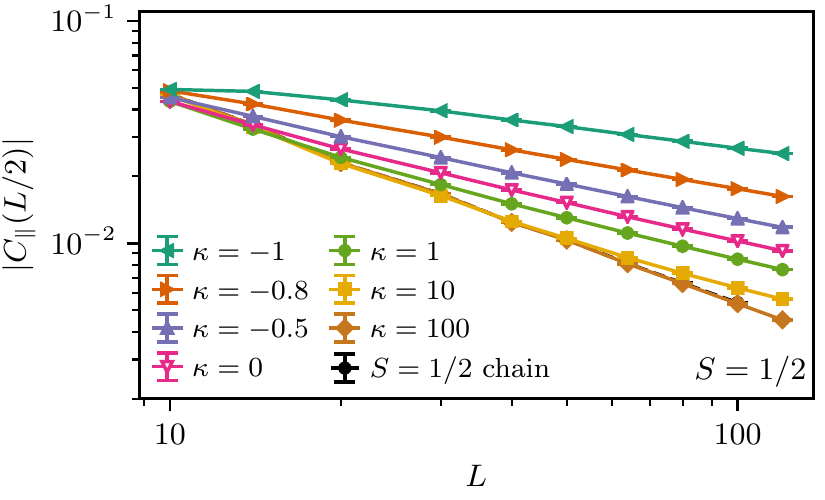}
\includegraphics{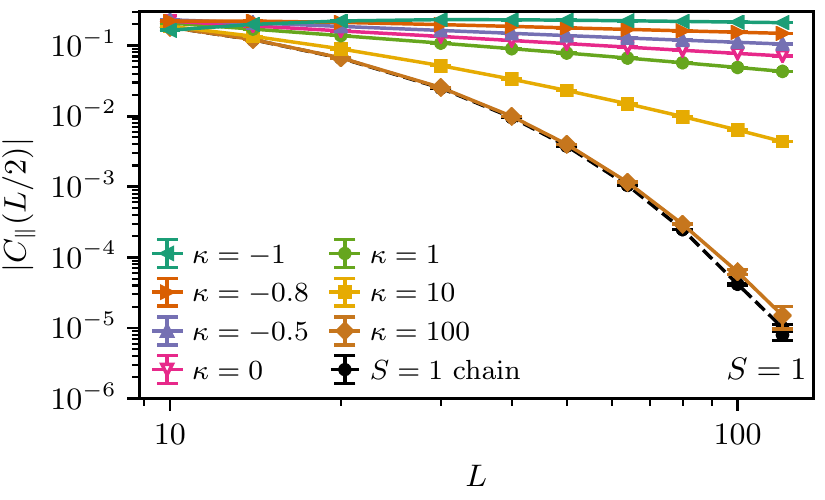}
\caption{Intra-edge correlations  $C_\parallel(L/2)$ as functions of $L$ for the perturbation shown in Fig.~\ref{fig_perturbations}(a) for the cases of $S=1/2$ (top panel) and $S=1$ (bottom panel).}
\label{fig_perturbation_a}
\end{figure}

The  finding above poses the question, to what extent the nonordinary  behavior is actually stable with respect to  perturbations being applied to  the edge-spin system. 
For example, in the case of classical surface criticality, different surface phases (if they are realized) can be accessed by varying the couplings at the surface, e.g., in the form of an enhanced coupling on the surface as compared to the bulk coupling strength. 
Similarly, we consider here a modified system~\cite{Weber18}, in which we enhance those exchange couplings that directly connect two nearest-neighbor dangling edge spins by a factor of $(1+\kappa)$ 
such that the case $\kappa=0$ corresponds to the original model. This modified setup is illustrated in Fig.~\ref{fig_perturbations}(a). For increasingly large values of $\kappa$, we  expect to eventually observe among the edge spins the correlations of the  limiting one-dimensional  spin chain. These are well known to be rather different for the $S=1/2$ and the $S=1$ cases: Whereas the former is dominated by a power-law decay, reflecting the gapless nature of the $S=1/2$ Heisenberg chain, the $S=1$ chain is characterized by an exponential decay, corresponding to its finite excitation gap. 
Based on the numerical data for the intra-edge correlations $C_\parallel(L/2)$, shown in  Fig.~\ref{fig_perturbation_a} for the accessible system sizes, we 
find that these distinct limiting correlations prevail on distances below a crossover length scale, which increase for larger values of $\kappa$. 
The QMC data shown in  Fig.~\ref{fig_perturbation_a}, furthermore, indicate that within the accessible system sizes the effective scaling exponent $\eta_\parallel$ slightly varies upon tuning the inter-edge coupling strength about $\kappa=0$. A possible implication is that the nonordinary critical exponents for the unperturbed system ($\kappa=0$) may be less universal than previously anticipated~\cite{Ding18} or that significantly larger system sizes are needed in order to probe the asymptotic scaling behavior, in particular, for nonzero values of $\kappa$. 
 
\begin{figure}[t]
\includegraphics{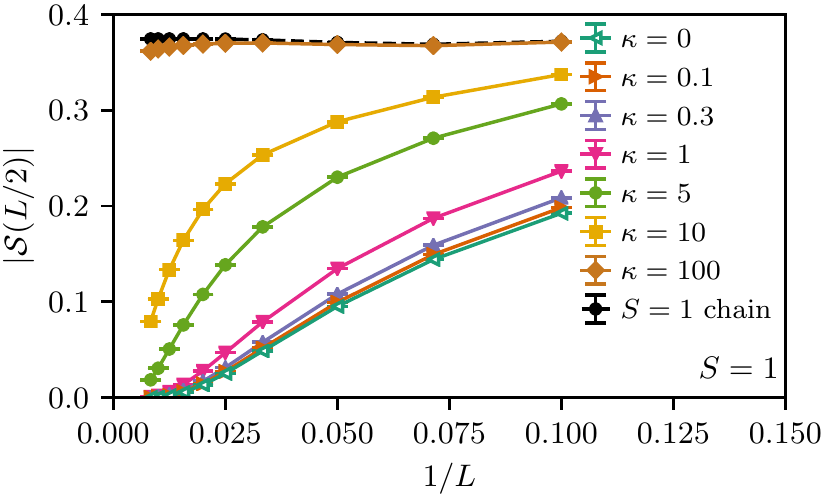}
\caption{String order correlations   $\mathcal{S}(L/2)$ as functions of $1/L$ for the perturbation shown in Fig.~\ref{fig_perturbations}(a) for the  $S=1$ system. Also shown are the results for an  isolated one-dimensional spin-1 Heisenberg chain.}
\label{fig_perturbation_a_so}
\end{figure}

We  note here that the nonlocal string order parameter~\cite{denNijs89}, which characterizes the symmetry-protected topological order~\cite{topo1,topo2} of the Haldane phase for the spin-1 Heisenberg chain, is  not stable with respect to the coupling of the edge-spin chain to  the bulk system. We  observe this breakdown of the string order parameter already within the quantum disordered regime, i.e.,  for even weaker values of $J/J_D$ than its quantum critical value: As shown in Appendix~\ref{appb}, the dangling spin-1 edge spins or $J/J_D=0.1$, indeed, exhibit the characteristic spin-spin correlations of the isolated spin-1 chain, whereas the nonlocal string order correlations exhibit a strong decay, indicating the absence of this string order along the edge spin-1 chain, due to the coupling of the edge spins to the bulk fluctuations. Similarly, we find that at the quantum critical point, the string order parameter is strongly suppressed, as shown in Fig.~\ref{fig_perturbation_a_so}.
 Here, we quantify  the nonlocal string order in terms of the corresponding correlation function at the maximum available distance 
$\mathcal{S}(L/2)=\langle S^z_1 \exp{[i\pi \sum_{j=2}^{L/2-1} S^z_j]} S^z_{L/2}\rangle$ where the index on a spin labels its position along the (edge) spin chain.  For an isolated spin-1 chain, this quantity converges to a  finite value at long chain lengths $L$. However, we observe that, for the as-cut dangling edge spins ($\kappa=0$), the
 string order correlations are strongly suppressed at large $L$. Whereas its magnitude increases upon increasing the intra-edge coupling,
 we find that, even for the largest considered values of $\kappa$,   eventually $\mathcal{S}(L/2)$ is  still suppressed at sufficiently  long length scales. As for the spin correlations, we observe only a gradual crossover to the limiting one-dimensional behavior. Beyond a corresponding length scale, which increases with increasing $\kappa$, the edge-spin subsystem, thus, retains its surface character. 
It might be interesting to explore if for finite edge chains---i.e., for open boundary conditions along both square lattice directions---the edge modes of the spin-1 Haldane phase, nevertheless, persist the bulk coupling, similar to the recently considered quantum critical one-dimensional spin-1 XXZ chain~\cite{Verresen19}. If this would be  the case, then the strong decay of the string order parameter would not be an appropriate indicator for the presence of the edge modes. One could, for example, try to address this question using
local spectroscopic measurements. This  extends, however, well beyond the scope of the present investigation, which  concerns  the static critical scaling properties of the critical edge-spin system.

\begin{figure}[t]
\includegraphics{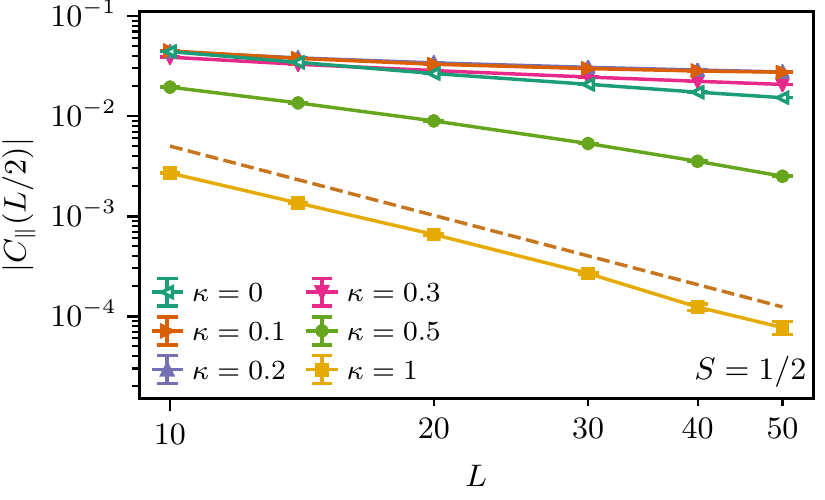}
\includegraphics{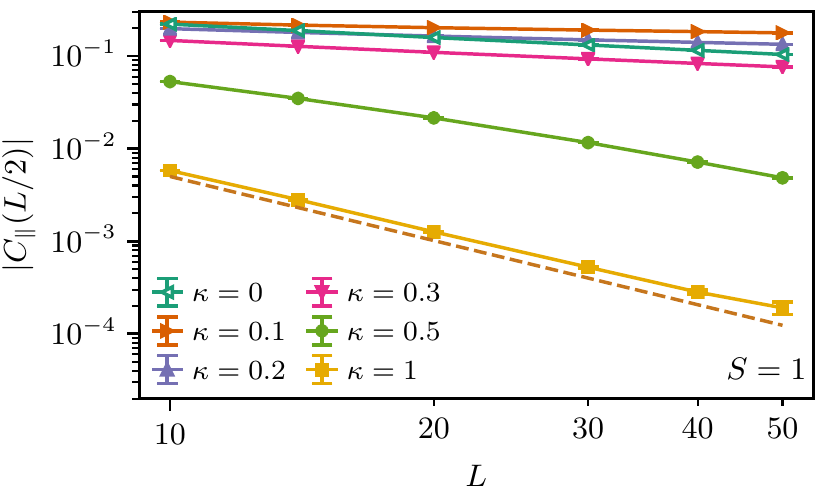}
\caption{  Intraedge correlations  $C_\parallel(L/2)$ as  functions of $L$ for the perturbation shown in Fig.~\ref{fig_perturbations}(b) for the cases of $S=1/2$ (top panel) and $S=1$ (bottom panel). In both panels, the dashed lines indicate a power-law decay corresponding to the ordinary case of the nondangling edge.}
\label{fig_perturbation_b}
\end{figure}

Finally, we note that, for negative values of $\kappa$, i.e., for a suppressed direct coupling along the edge, the intra-edge-spin correlations are, in fact,  {\it enhanced}  compared to the case of $\kappa=0$  for both  $S=1/2$ and $S=1$. 
In particular, note that, for $\kappa=-1$, the direct coupling along neighboring edge spins is completely removed. The rather strongly enhanced correlations in this limit suggest that the finite inter-edge coupling competes to some extent with the antiferromagnetic correlations induced by the critical bulk fluctuations.  One possible reason would be an enhanced tendency of the intraedge coupling to promote the formation of local singlet fluctuations.  Such a
genuine quantum effect   is not captured within the corresponding classical surface criticality. 
It would, thus, be  interesting to characterize the underlying mechanism in terms of an effective edge-only model that results from integrating out the bulk degrees of freedom, inducing thereby an effective (retarded) interaction among the edge spins, which we expect to be long-ranged due to the bulk criticality. Such an analysis extends, however, well beyond the scope of the present investigation. 

There is  a further  means of probing  the nature of the nonordinary edge-spin correlations within our numerical approach. In particular,  it is feasible to introduce a perturbation to the edge-spin system that allows us to continuously tune between the dangling and the nondangling cases~\cite{Weber18}. For this purpose, we  introduce a new row of spins and couple it to the original edge spins through a coupling of strength $\kappa J_D$, as shown in Fig.~\ref{fig_perturbations}(b).
Since small values of $\kappa$ lead to a weak additional coupling as compared to the original couplings, the temperature scaling has been adapted to $T = \kappa_\mathrm{min}/(2L)$ for these simulations, where $\kappa_\mathrm{min}=0.1$ is the smallest finite value of $\kappa$ that we considered. In this way, we ensure that we still probe ground state correlations, i.e., we assure that the effect of the new coupling is not suppressed by thermal fluctuations. This adaptation, however, increases the computational cost, and it, thus, limits the accessible system sizes. 

The  results from the QMC simulations on these perturbed edges are shown in 
Fig.~\ref{fig_perturbation_b}. We  again observe qualitatively similar behavior for both values of spin $S$: For weak values of $\kappa$, the correlations along the edge spins are enhanced as compared to the case of $\kappa=0$. This shows that the additional coupled spins initially enhance the antiferromagnetic correlations along the edges.  However, a further increase in $\kappa$  leads to the suppression of the correlations, which  eventually exhibit for $\kappa=1$ a spatial decay that is, indeed, in  accord with the scaling  observed for the ordinary case (which is indicated in Fig.~\ref{fig_perturbation_b} by the dashed lines).  We, thus, find that, for this edge perturbation, which preserves 
 the quasi-two-dimensional character of the edge-spin system, both $S=1/2$ and $S=1$ continue to exhibit their very similar behavior. 

\section{Dimer MF theory}
\label{sec_mf}
\begin{figure}[t]
\includegraphics[width=0.7\columnwidth]{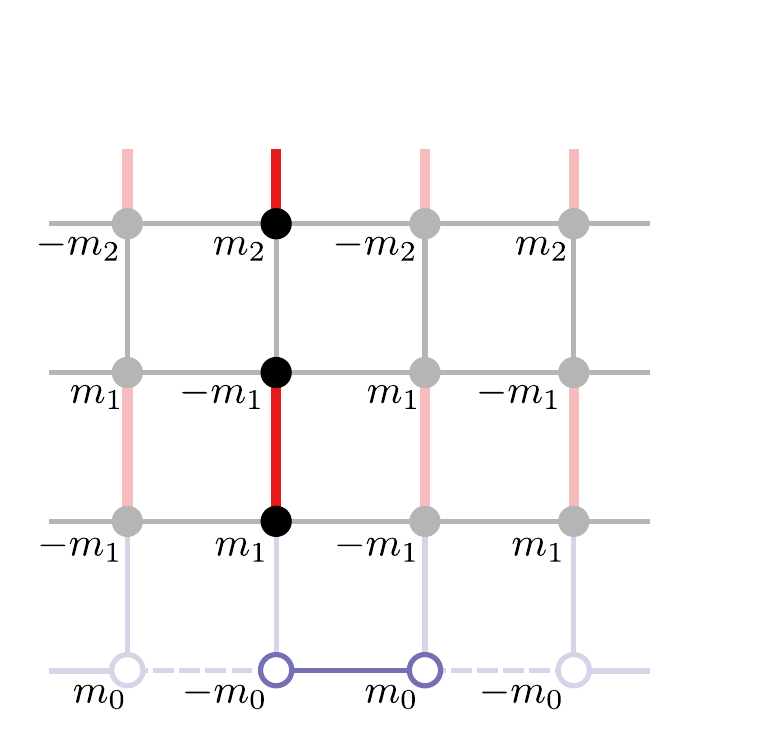}
\caption{  Structure of the MF decoupling of the columnar-dimer lattice model for the case of  dangling edge spins. The lattice is considered semi-infinite in the vertical ($y$) direction, and has a horizontal extent that is specified by the number $L$ of edge spins ($L=4$ in the figure), where periodic boundary conditions are considered. The opaque dimers are part of the MF Hamiltonian. The remaining degrees of freedom are given by translation symmetry. The bottom row shows the special construction to treat the dangling edge-spin condition.}
\label{fig_mf_lattice}
\end{figure}

A simple means of rationalizing the similar qualitative behavior at the edges of the columnar-dimer lattice for both $S=1/2$ and $S=1$ can be obtained  from an appropriate MF theory, as detailed in this section. Although we do not expect to be able to quantitatively describe within such a MF approach the quantum critical scaling exponents, we instead aim for  a simple analytic account that captures the qualitative features observed in the QMC simulations. 
For this purpose, we calculated the magnetization profile of a semi-infinite spin-$S$ columnar-dimer system using a dimer-based cluster MF theory cf. Fig.~\ref{fig_mf_lattice}. Within this approach, we examine the system in the  antiferromagnetic regime near the quantum critical point where the bulk order is weak. As shown below, the system is, indeed, found to  exhibit rather distinct magnetization profiles, depending on whether the edge spins are dangling or nondangling. Moreover, this qualitative distinction appears irrespective of  the value of the spin quantum number $S$, in accord with the QMC findings. The MF theory formulated below shares some similarity to the bond-operator MF approach for the $S=1/2$ case~\cite{Sachdev90, Fritz11} and can readily be formulated for higher values of $S$.

\subsection{MF equations}

We start from the Hamiltonian in Eq.~\eqref{eq_hamiltonian} and fix $J_D=1$ for convenience in this section. 
To account for the dominant singlet formation on the dimer ($J_D$) bonds, which leads to the presence of the quantum critical point, we perform a MF  decoupling of the interdimer coupling ($J$), whereas leaving the dimer bonds intact. Since the system is translationally  invariant parallel to its edge, we can label the magnetic MF parameters in terms of the perpendicular distance from the edge. More specifically, we denote the distance of the center of a dimer from the edge by $y$ and refer to the two spins of that dimer as $\mathbf{S}_{y,1}$ and  $\mathbf{S}_{y,2}$ where the former is the one closer to the edge. We, then, define the MF parameter $m_y$ in terms of the ground-state expectation values $\langle S^z_{y,1}\rangle$ and $\langle S^z_{y,2}\rangle$. Close to the quantum critical point, these are small compared to $S$, and
since the ground state of a single quantum spin dimer resides in the zero magnetization sector, we obtain 
\begin{equation}\label{eq:mz}
m_y=\langle S^z_{y,1}\rangle=-\langle S^z_{y,2}\rangle.
\end{equation}

Applying the decoupling of the interdimer couplings  results in the following MF Hamiltonian for a semi-infinite system with $L$  edge spins (cf.~Fig.~\ref{fig_mf_lattice}): 
\begin{align}
\label{eq:hmf}
H_\text{MF} &= L \Big(\sum_{y=1}^\infty \mathbf{S}_{y,1} \cdot \mathbf{S}_{y,2} - J \left(2 m_y + m_{y-1}\right) S^z_{y,1}\nonumber
\\
&+ J \left(2 m_y + m_{y+1}\right)S^z_{y,2} \Big)
- J \sum_{\langle i,j\rangle}{} \langle S^z_i\rangle\langle S^z_j\rangle,
\end{align}
where the first summation extends over all the dimers within a single column of the semi-infinite system, and the second summation extends over all  interdimer ($J$) bonds of the lattice. 
The MF solution minimizes $\langle H_\text{MF}\rangle$ under the condition in Eq.~\eqref{eq:mz}. One can express $\langle H_\text{MF}\rangle$ and the resulting MF  equations explicitly in terms of the
ground-state energy $E_S[h_1,h_2]=\langle H_S\rangle$ and the local magnetization $M_S[h_1,h_2] = \langle S_1^z\rangle$  of a single spin-$S$ dimer, with the Hamiltonian,
\begin{equation}
H_S = \mathbf{S}_1\cdot \mathbf{S}_2 - h_1 S_1^z - h_2 S_2^z,
\end{equation}
which can be  readily calculated exactly. In terms of these functions, we obtain
\begin{align}
\frac{\langle H_\text{MF}\rangle}{L} &= \sum_{y=1}^\infty E_S[J(2 m_y + m_{y-1}), -J(2 m_y + m_{y+1})]\nonumber\\ &+ J \sum_{y=1}^\infty m_y (2m_y+m_{y+1}),
\end{align}
and
\begin{align}
\label{eq:mbulkmf}
m_y &= M_S[J(2 m_y + m_{y-1}), -J(2 m_y + m_{y+1})].
\end{align}
This self-consistency equation can be  solved numerically,  given appropriate boundary conditions at $y=0$ and for $y\rightarrow\infty$, to be considered next. 

\subsection{Boundary conditions}
For $y\rightarrow\infty$, the value of  $m_y$ approaches the  bulk magnetization of the infinite  model, denoted by $m_b$. From the self-consistency condition, $m_b$ is then determined through
\begin{equation}
\label{eq:bulkboundary}
m_b = M_S[3 J m_b, -3 J m_b].
\end{equation}
The boundary condition at $y=0$ depends on the details of the  edge configuration. 
The nondangling case can be implemented by setting $m_0 = 0$ so that effectively the nondangling edge is formed by the bottom row of dimers.

The case of dangling edge spins is more complicated to capture since the current MF approach cannot treat unpaired spins. Hence, we introduce an artificial grouping of the edge spins into dimers (cf.~Fig.~\ref{fig_mf_lattice}), which allows us  to use the dimer MF decoupling scheme also for the dangling edge spins. Note that, in terms of the exchange couplings, translational symmetry is preserved by this grouping of the edge spins. 
For comparison with the QMC results, we, furthermore, allow for a modified  coupling $J (1+\kappa)$ among neighboring edge spins and then arrive  at the following self-consistency equation, which determines the MF parameter $m_0$  for the dangling case:
\begin{equation}
m_0 = M_S\!\Big[m_0 + \frac{m_1}{1+\kappa},- m_0 - \frac{m_1}{1+\kappa}\Big].
\end{equation}
This equation  can be solved independently of the bulk to yield a function $m_0=m_0(m_1)$, considering $m_1$ as an external parameter. Note, that this equation does not  depend on the coupling parameter $J$, and $\kappa$ merely renormalizes the external field $m_1/(1+\kappa)$. As such, the edge turns out to be always ordered because we find
\begin{equation}
\frac{d M_S[h,-h]}{dh}\Big\vert_{h=0} \ge 1,
\end{equation}
for all values of spin $S$. 
In the full treatment of the model, fluctuations will destroy this surface order.
However, its presence for the dangling case within this MF approach corresponds nicely to the strongly enhanced intra-edge--spin-spin correlations observed in the  QMC study of  dangling edge spins, as compared to the nondangling case.

\subsection{Connection to the continuum model}
\begin{figure}[t]
\includegraphics{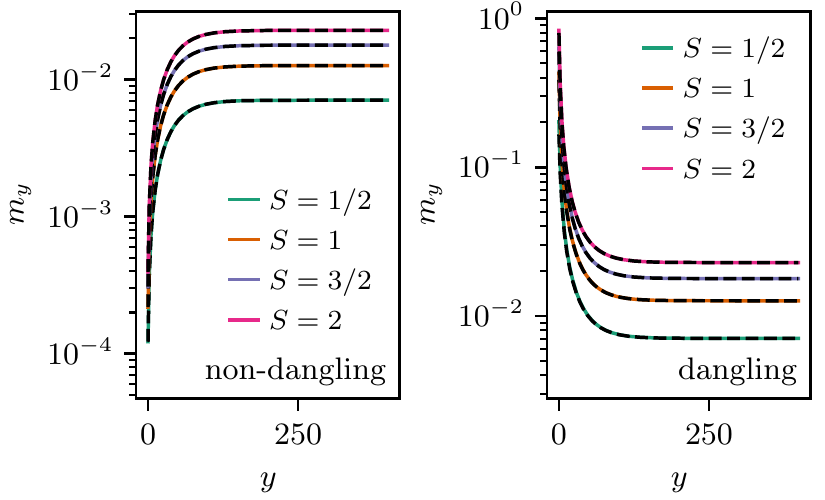}
\caption{  Comparison between the numerical solution of  the lattice model  MF theory in Eq.~\eqref{eq:mbulkmf} (solid lines) to the analytical MF theory of the O($n$) model (dashed lines)  for $\tau = -6\times 10^{-4}$ and $L=400$ for different  values of spin $S$. The left (right) panel  shows the nondangling (dangling) edge-spin case.}
\label{fig_meanfield}
\end{figure}

Close to the quantum critical point, $m_y\ll S$ is small, and we can approximate
\begin{equation}
M_S[h,-h] = a_S h - b_S h^3 + \mathcal{O}\!\left(h^5\right)
\end{equation}
where the parameters $a_S$ and $b_S$ depend on $S$, and can be calculated from solving the spin-$S$ single dimer system. Substituting in Eq.~\eqref{eq:mbulkmf} and expanding
\begin{equation}
m_{y\pm 1} \approx m(y) \pm \partial_y m(y) + \frac 1 2 \partial_y^2 m(y),
\end{equation}
leads to
\begin{equation}
\left(-\partial_y^2 + \frac{2}{a_SJ} - 6 + \frac{2 \times 3^3 \:b_S J^2}{a_S} m(y)^2\right) m(y) = 0,
\end{equation}
if we discard higher-order terms, such as  $m(y)^2\partial^2_y m(y)$. Defining new variables,
\begin{align}
\tau &= \frac{2}{a_SJ} - 6\\
u &= \frac{4\times 3^4 \: b_S J^2}{a_S} \stackrel{|\tau| \ll 1}{\approx} \frac{36 \:b_S}{a_S^3}
\end{align}
we obtain a differential equation
\begin{equation}
\left(-\partial^2_y + \tau + \frac{u}{3!} m(y)^2\right) m(y) = 0,
\label{eq:contmf}
\end{equation}
which is exactly the MF equation of the continuous semi-infinite O($N$) model~\cite{Lubensky75, Diehl86}. We note that, within our approach, 
the normalization of $m(y)$ is fixed to the asymptotic bulk value of the order parameters in the MF theory of the original lattice model (cf. also Fig.~\ref{fig_meanfield}).
In terms of $\tau$, the bulk system resides within the disordered (ordered) phase for $\tau>0$ ($\tau<0$), with the bulk critical point located at $\tau=0$.
From this continuum form, we can, thus, directly read off the value of the bulk critical coupling of the columnar-dimer model within our dimer-MF approach,
\begin{equation}
J_c^\text{MF} = \frac{1}{3a_S} =\begin{cases} \frac 1 3, & S=\frac 1 2\\ \frac 1 8, & S=1\end{cases},
\end{equation}
given in units of $J_D$. For the $S=1/2$ case, the  above value agrees with the critical coupling ratio reported in Ref.~\onlinecite{Fritz11}, obtained within the bond-operator MF theory for the columnar-dimer lattice. 
For the further analysis of the magnetization profile within this continuum description, we need to again consider the boundary condition of the magnetization profile $m(y)$  near the edge. As  mentioned already at the beginning of this section, we perform our calculations within the regime of finite, weak bulk order, corresponding to $\tau<0$. 

Within the classical MF theory of surface critical phenomena, the boundary condition for the magnetization profile can then be specified in terms of   
the surface enhancement  parameter $c$, which,  within MF theory, is given in terms of the slope of the magnetization profile near the boundary,
\begin{equation}
\label{eq:cont_boundary}
\frac{\text{d}}{\text{d}y} m(y) \Big\vert_{y=0} = c\,m(0).
\end{equation}

At the MF level, its value distinguishes three different surface transitions, namely,
$c>0$ leads to the ordinary transition, $c=0$ leads to the special, and 
$c<0$ leads to the extraordinary transition, respectively.

To draw a connection to the lattice boundary conditions, we  approximate the slope at the boundary by the discrete derivative 
\begin{align}
\frac{\text{d}}{\text{d}y} m(y) \Big\vert_{y=0} &\approx m_1 - m_0(m_1) \nonumber\\
&\approx m(1,c,\tau,u)-m_0(m(1,c,\tau,u)),
\end{align}
in terms of  the  solution $m(y,c,\tau,u)$ of the continuum equation. In the nondangling case we can use $m_0(m_1) = 0$ to estimate the boundary slope in $m(y)$ for small $y$ and obtain  $c\approx 1$. In the dangling case, we obtain an implicit equation for $c$ that gives rise to a solution with $c<0$.
In fact, the analytical continuum solution $m(y,c,\tau,u)$ with the model parameters determined as described above  matches the numerical solution of Eq.~\eqref{eq:mbulkmf} rather well (cf.~Fig.~\ref{fig_meanfield}). We find that, for the nondangling case, the magnetization profile decreases upon approaching the edge, whereas for the dangling case, it increases instead. This qualitative distinction is, furthermore, observed irrespective of the actual value of $S$. 

In view of the approximative character of our MF approach, we  draw the following conclusions from the  analysis of this section: (i) Within the MF approximation, we observe distinctly different behavior for the case of nondangling and dangling edge spins. 
(ii) This distinction results irrespective of  the spin value $S$, as observed also in the QMC simulations. 
(iii) The dimer MF theory provides a formal mapping to the continuum MF theory of the corresponding classical surface criticality of the three-dimensional O($N$) model. In particular, for the nondangling case, the corresponding classical surface criticality, in fact, belongs to the ordinary case. 
(iv) The  enhanced correlations for dangling edge spins observed in the QMC simulations relate to an ordered edge within the MF approximation. This is  a generic property of the nondangling case, even in the presence of a modified edge coupling strength.  (v)  Within our MF approach, the effective continuum description for the dangling case falls within the extended regime of the extraordinary surface criticality. This is a MF artifact  however, because  quantum fluctuations  destroy the surface order along the one-dimensional edge as observed in the QMC calculations.  
\section{Conclusions}
\label{sec_conclusions}
Based on unbiased quantum Monte Carlo simulations, we demonstrated that 
dangling edge spins of two-dimensional quantum critical edge spins exhibit
nonordinary correlations irrespective of the values of spin $S=1$ or $S=1/2$.
Focusing on the columnar-dimer lattice, we found that the distinction from 
the nondangling case, which exhibits ordinary scaling exponents, is, hence, not in
direct correspondence to the qualitatively distinct behavior of the corresponding single chain physics. 
These findings  indicate that  attempts to link the emergence of such nonordinary edge criticality 
to Berry phase effects via a topological $\theta$ term in the low-energy effective action for 
the one-dimensional limit of  dangling spin-1/2 spins cannot 
account appropriately for this unconventional scaling behavior. Finally, we presented a cluster mean-field theory that exposes the 
edge-spin configuration (dangling vs nondangling) as the  relevant characteristics to observe
the nonordinary surface criticality irrespective of the quantum spin number $S$. 
We hope that our results  motivate the development of refined analytical 
treatments of the quantum fluctuations in these quantum critical edge-spin systems. This may provide a quantitative understanding
on the observed peculiar edge criticality and the corresponding scaling exponents.

\begin{acknowledgments}
We thank L. Fritz, F. Parisen Toldin, and F. Pollmann for discussions and 
acknowledge support by the Deutsche Forschungsgemeinschaft through Grant No. WE/3649/4-2 of the FOR 1807 and through RTG 1995. Furthermore, we thank the IT Center at RWTH Aachen University and the JSC J\"ulich for access to computing time through JARA-HPC.
\end{acknowledgments}

\appendix

\section{Bulk Critical Point for $S=1$}
\label{appa}
We determined the quantum critical point for the $S=1$ Heisenberg model on the columnar-dimer lattice using the crossing method on observables with known critical finite-size scaling~\cite{Wang06}.
For this purpose, 
we simulated systems of $L\times L$ two-spin dimer unit cells, with a total of $N=2 L^2$ spins using periodic boundary conditions. The temperature was scaled as $T=J_D/(2L)$, since the dynamical critical exponent for the bulk transition is $z=1$ in order to probe the ground-state properties. 
We measured the Binder ratio $Q = \langle m_s^4\rangle/\langle m_s^2\rangle^2$ 
of the staggered  bulk magnetization $m_s=1/N\sum_i \varepsilon_i S^z_i$. Here, $\epsilon_i=\pm 1$, depending on the sublattice to which site $i$ belongs. In addition, we measured
the uniform susceptibility $\chi_u=\beta/N\,\langle (\sum_i S^z_i )^2\rangle$ of the bulk system. 

\begin{figure}[t]
\includegraphics{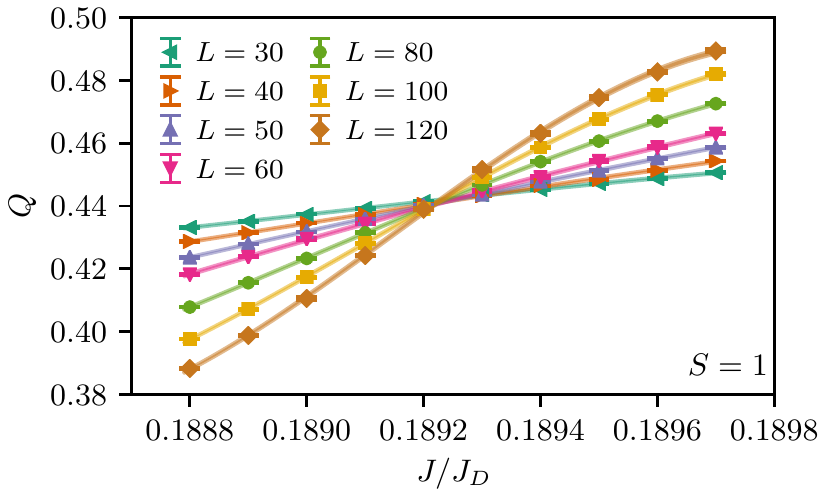}
\caption{Binder ratio $Q$ of the spin-1 Heisenberg model on the columnar-dimer lattice for different system sizes $L$ as a function of  $J/J_D$. Close to the crossing point, the data has been approximated by polynomials of degree $d$ (shown here for $d=3$).}
\label{fig_bulkcrit_binder}
\end{figure}
\begin{figure}[t]
\includegraphics{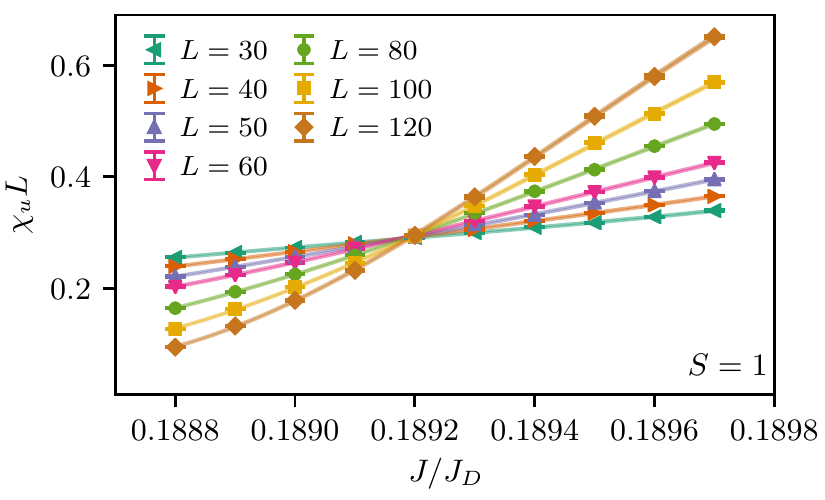}
\caption{Uniform susceptibility $\chi_u$ multiplied by system size $L$ of the spin-1 Heisenberg model on the columnar-dimer lattice for different $L$’s as a function of  $J/J_D$. Close to the crossing point, they have been approximated by polynomials of degree $d$ (shown here for $d=3$).} 
\label{fig_bulkcrit_chiu}
\end{figure}
%
\begin{figure}[t]
\includegraphics{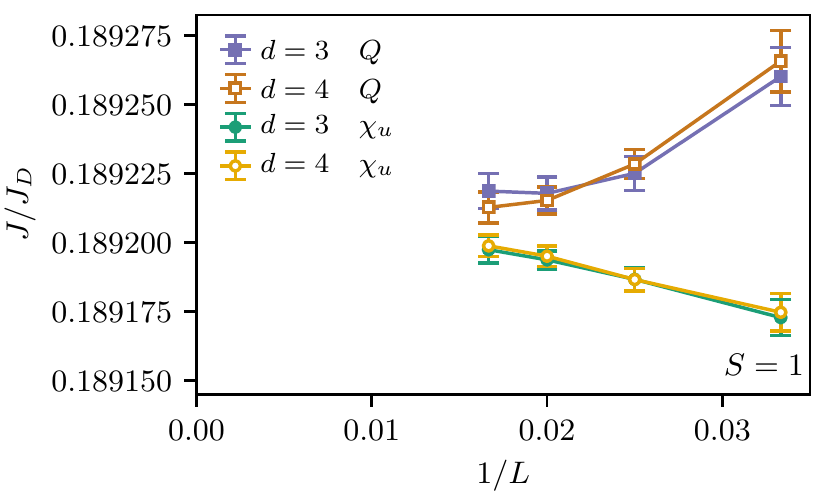}
\caption{  Crossing points in the Binder ratio $Q$ and the uniform susceptibility $\chi_u L$ between the fitting polynomials for systems with linear sizes $L$ and $2L$ vs the  inverse linear system size $1/L$ for the spin-1 Heisenberg model on the columnar-dimer lattice. The fitting polynomial has degree $d$, and the range of system sizes is $L=30, 40, 50, 60$. The connecting lines are a guide for the eye.}
\label{fig_bulkcrit_scaling}
\end{figure}

The results for different system sizes are shown in Figs.~\ref{fig_bulkcrit_binder} and ~\ref{fig_bulkcrit_chiu}, respectively. To obtain a more accurate estimate of the crossing point, we fitted a polynomial of degree $d$ to the data around the critical point. 
The broadening of the curves shows the bootstrap error of the fitting procedure. To examine the finite size effects, we collected the crossing positions of the finite-size data for different system sizes in Fig.~\ref{fig_bulkcrit_scaling} and fitting polynomials of degrees $d=3$ and $d=4$. The crossings of $Q$ and $\chi_u L$  converge from opposite directions, thus bracketing a common limiting value. 
The interpolations using 
$d=3$ and $d=4$ lead to consistent results within our statistical accuracy, and we, thus,  arrive at a final estimate of $J_c/J_D = 0.189\,20(2)$.

\section{Disordered Bulk for $S=1$}
\label{appb}
\begin{figure}[ht]
\includegraphics{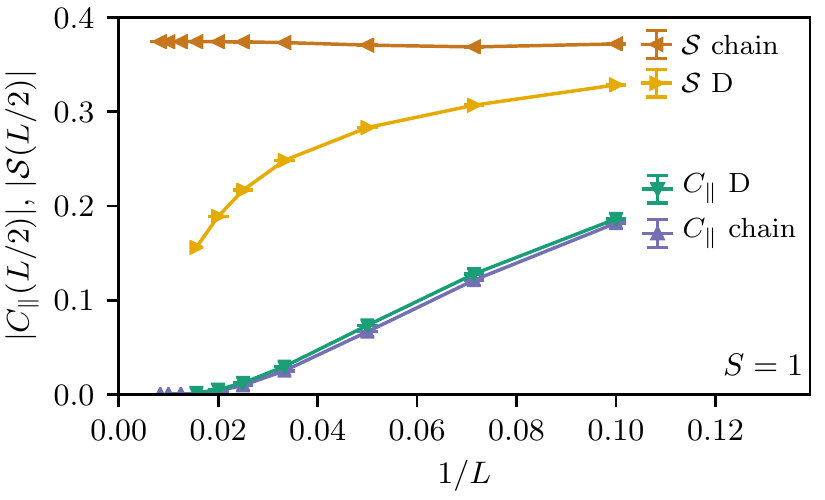}
\caption{  Spin-correlation function $C_\parallel(L/2)$ and string order  correlation function $\mathcal{S}(L/2)$ along the dangling edge spins on the spin-1 Heisenberg model on the  columnar-dimer lattice (denoted  as D) at a values of $J/J_D=0.1$ inside the bulk-disordered regime and for an isolated one-dimensional spin-1 Heisenberg chain for comparison (denoted as chain) vs the  inverse linear system size $1/L$. For these QMC simulations, $T=J/(2L)$ was used, and lines are guides to the eye.}
\label{fig_surfnoncrit}
\end{figure}

Within the  bulk-disordered phase at  $J/J_D=0.1$, the finite-range bulk correlations only weakly affect the correlation function between distant edge spins, which, thus, closely resemble the behavior of the one-dimensional spin-1 Heisenberg chain (cf.~Fig.~\ref{fig_surfnoncrit}). On the other hand, the nonlocal string order~\cite{denNijs89}, which characterizes the Haldane phase of the one-dimensional spin-1  chain,  appears to be  unstable against the bulk coupling as seen from Fig.~\ref{fig_surfnoncrit}: Here, we quantify  the string order in terms of the corresponding correlation function at the maximum available distance, 
$\mathcal{S}(L/2)=\langle S^z_1 \exp{[i\pi \sum_{j=2}^{L/2-1} S^z_j]} S^z_{L/2}\rangle$, where the index on a spin labels its position along the edge-spin chain.  For the isolated spin-1 chain, this quantity converges to a  finite value for long chain lengths $L$. 
Our QMC data  for the $S=1$ columnar-dimer lattice with dangling edge spins at $J/J_D=0.1$ display, instead, a steady  suppression of $\mathcal{S}(L/2)$ with increasing system size, indicating that the string order parameter vanishes in the thermodynamic limit already within the  bulk-disordered regime. As discussed in the main text, this suppression is even more pronounced at the quantum critical point.

%
%

\begin{references}



\bibitem{Sachdev11} S. Sachdev, {\it Quantum Phase Transitions} (Cambridge University Press, Cambridge, 2011).

\bibitem{Diehl86} H. W. Diehl, in {\it Phase Transitions and Critical Phenomena}, edited by C. Domb and J. L. Lebowitz (Academic, London, 1986), Vol. 10.


\bibitem{Zhang17} L. Zhang and F. Wang, Phys. Rev. Lett. {\bf 118}, 087201 (2017).


\bibitem{Ding18} C. Ding, L. Zhang, and W. Guo, Phys. Rev. Lett. {\bf 120}, 235701 (2018).


\bibitem{Weber18} L. Weber, F. Parisen Toldin, and S. Wessel, Phys. Rev. B {\bf 98}, 140403(R) (2018).


\bibitem{Mermin66} N. D. Mermin and H. Wagner, Phys. Rev. Lett. {\bf 17}, 1133 (1966).


\bibitem{Matsumoto01} M. Matsumoto, C. Yasuda, S. Todo, and H. Takayama, Phys. Rev. B {\bf 65}, 014407 (2001).


\bibitem{Wenzel08} S. Wenzel, L. Bogacz, and W. Janke, Phys. Rev. Lett. {\bf 101}, 127202 (2008).


\bibitem{Haldane81} F. D. M. Haldane, ILL  Report No. SP-81/95, 1981 (unpublished).


\bibitem{Haldane83a}  F. D. M. Haldane,  Phys. Rev. Lett. 50, 1153 (1983).


\bibitem{Haldane83b} F. D. M. Haldane, Phys. Lett. {\bf 93A}, 464 (1983).


\bibitem{Affleck85a} I. Affleck, Nucl. Phys. B 257, 397 (1985).


\bibitem{Affleck85b} I. Affleck, Nucl. Phys. B 265, 409 (1985).


\bibitem{Haldane85} F. D. M. Haldane, J. Appl. Phys. {\bf 57}, 3359 (1985).


\bibitem{Sandvik91} A. W. Sandvik and J. Kurkijärvi, Phys. Rev. B {\bf 43}, 5950 (1991).


\bibitem{Sandvik99} A. W. Sandvik, Phys. Rev. B {\bf 59}, 14157(R) (1999).


\bibitem{Henelius00} P. Henelius, A.W. Sandvik, Phys. Rev. B 62, 1102 (2000). 

\bibitem{Campostrini02} M. Campostrini, M. Hasenbusch, A. Pelissetto, P. Rossi, E. Vicari, Phys. Rev. B {\bf 65}, 144520 (2002).

\bibitem{denNijs89} M. den Nijs and K. Rommelse, Phys. Rev. B {\bf 40}, 4709 (1989).


\bibitem{topo1} Z.-C. Gu and X.-G. Wen, Phys. Rev. B {\bf 80}, 155131 (2009).


\bibitem{topo2} F. Pollmann, A. M. Turner, E. Berg, and M. Oshikawa, Phys. Rev. B {\bf 81}, 064439 (2010).


\bibitem{Verresen19} R. Verresen, R. Thorngren, N. G. Jones, and F. Pollmann, arXiv:1905.06969.


\bibitem{Sachdev90} S. Sachdev and R. N. Bhatt, Phys. Rev. B {\bf 41}, 9323 (1990).

\bibitem{Fritz11} L. Fritz, R. L. Doretto, S. Wessel, S. Wenzel, S. Burdin, and M. Vojta, Phys. Rev. B {\bf 83}, 174416 (2011).

\bibitem{Lubensky75} T. C. Lubensky, M. H. Rubin, Phys. Rev. B {\bf 12}, 3885 (1975).


\bibitem{Wang06} L. Wang, K. S. D. Beach, and A. W. Sandvik, Phys. Rev. B {\bf 73}, 014431 (2006).


\end{references}
\end{document}